\documentclass[useAMS,usenatbib,usegraphicx]{mn2e}

\newcommand{\kms}{${\rm km~s}^{-1}$}
\usepackage{amssymb}
\begin{document}

\title[Detection of Satellite Remnants: mGC3 Method]
{Detection of Satellite Remnants in the Galactic Halo with Gaia -- II. 
A modified Great Circle Cell Method}

\author[C. Mateu et al.]
{C.~Mateu,$^1$\thanks{cmateu@cida.ve} G.~Bruzual,$^1$ L.~Aguilar,$^2$ A.G.A~Brown,$^3$ 
\newauthor
O. Valenzuela,$^4$ L.~Carigi,$^4$ F.~Hern\'andez,$^1$ H.~Vel\'azquez $^2$  \\
$^{1}${{Centro de Investigaciones de Astronom\'{\i}a, AP 264, M\'erida 5101-A, Venezuela}}\\
$^{2}${{Instituto de Astronomia, UNAM, Apartado Postal 877, 22860 Ensenada, B.C., M\'exico}}\\
$^{3}${{Sterrewacht Leiden, Leiden University, PO Box 9513, 2300 RA Leiden, the Netherlands}}\\
$^{4}${{Instituto de Astronom\'{\i}a, UNAM, Apartado Postal 70-264, 04510 M\'exico, D.F., M\'exico}}
}
\maketitle

\begin{abstract}
We propose an extension of the GC3 streamer finding method of \citet{joh96} that can be applied to the
future Gaia database. The original method looks for streamers along great circles in the sky, our extension
adds the kinematical restriction that velocity vectors should also be constrained to lie along these great circles, as seen by a Galactocentric observer. We show how to use these combined criteria starting from heliocentric observables. We test it by using the mock Gaia catalogue of \citet{bro05}, which includes a realistic Galactic background and observational errors, but with the addition of detailed star formation histories for the simulated satellites.
We investigate its success rate as a function of initial satellite luminosity, star formation history and orbit.
We find that the inclusion of the kinematical restriction vastly enhances the contrast between a streamer and
the background, even in the presence of observational errors, provided we use only data with good astrometric
quality (fractional errors of 30 per cent or better). The global nature of the method diminishes the erasing effect of
phase mixing and permits the recovery of merger events of reasonable dynamical age. Satellites with a
star formation history different to that of the Galactic background are also better isolated. We find that satellites
in the range of $10^8-10^9$ L$_\odot$ can be recovered even for events as old as $\sim10$ Gyr. Even satellites with $4-5\times10^7$ L$_\odot$ can be recovered for certain combinations of dynamical ages and orbits.

\end{abstract}

\nokeywords

\section{Introduction}\label{s:intro}

  The $\Lambda$ cold dark matter ($\Lambda$CDM) hierarchical paradigm is the best model we have to explain the large-scale structure of the Universe \citep*{Springel06, Spergel07,  Klypin2010}.  In recent years, it has proved
to successfully reproduce a number of observational measurements (e.g. spatial and colour-magnitude
distributions) for galaxies observed in the local Universe and at higher redshift (for  recent reviews 
see \citealt{Baugh06, avila-reese06} ). In spite of the encouraging progress on the large scale,  at galactic
and subgalactic scales, the success of the model has not been convincingly demonstrated as yet, and
a number of issues remain subject of a lively debate in the astronomical community. Examples of  these
are  the core--cusp issue   \citep*{dBlok2010, Governato10, Puglielli2010,Valenzuela07,Gentile04}, 
the missing satellites problem \citep{KKVP99,Moore99,Kravtsov2010}, the angular momentum problem 
\citep{Abadi03,Governato04,Okamoto05,Governato07}, or the Downsizing--Specific Star Formation Rate
relation for dwarf galaxies \citep*{Colin2010,Firmani09,Bauer05}. 
  
Another powerful test for the $\Lambda$CDM scenario at subgalactic scales,  may be provided by galaxy halos,
in particular by the Milky Way (MW) stellar halo.  If the Galaxy was assembled in a hierarchical way, there must be
fossil signatures  in the phase space distribution of halo stars and also in the Galactic disc, although kinematic
features in the latter may have multiple origins \citep{Antoja09, Minchev09,FGomez010}.  Theoretically, the origin
and structure of the stellar halo have been studied by several authors using a variety of techniques (for a review,
see Helmi 2008). These include cosmological numerical simulations with and without baryonic physics 
\citep*{Zolotov09,Diemand05,Abadi03} and phenomenological modelling of the evolution of baryons inside haloes, 
usually in combination with N-body simulations that provide the dynamical history of the system \citep{BJ05,Cooper09}.
Historically, chemical and kinematic information was used as a basis to formulate the first galaxy formation models.
In their classical paper, \citet*{ELS1962} suggested  a rapid radial collapse that later continued to form the stellar disc. 
About a decade later, \citet{SZ1978} formulated the hypothesis that the stellar halo formed over a longer time-scale, 
through the agglomeration of many subgalactic `fragments' that may be similar to the surviving dwarf spheroidal galaxies  observed today as satellites of the Milky Way. The  \citet{SZ1978} scenario and several  observational results  
are in qualitative agreement with expectations from the $\Lambda$CDM model \citep{DLucia08}. 
However, the observed abundance pattern \citep{Tolstoy09} seems to exclude the possibility that a significant contribution to the stellar 
halo comes from disrupted satellites similar to the present-day dSphs \citep{BJ05,Robertson05,Font06}.  

Over the next decade, a number of astrometric and spectroscopic surveys will provide accurate spatial, kinematic
and chemical information for a large number of stars e.g. the Gaia satellite \citep{Lindegren08}; the Radial Velocity 
Experiment-RAVE  \citep{RAVESteinmetz06} and the Sloan Extension for Galactic Understanding and Exploration-
SEGUE \citep{Yanny09}. This vast fossil records will provide important advances in our understanding of the 
sequence of events which led to the formation of our Galaxy \citep{Freeman-Bland02}.

It is critical to design efficient strategies in order to extract valuable dynamical information from the plethora of upcoming observations.  The task is not a trivial one. It is not clear which is the optimal  strategy, or in which space substructure can be best located.  Natural diagnostics like  integrals of motion  variables may be hampered by the finite accuracy of surveys \citep{Helmi99, hel00}. Recently, action-angle variables have shown to provide a promising avenue \citep{FGomez010}. In addition, \citet{Font06} showed the usefulness of  chemical information in identifying the relics of merger events. 

In the present work we continue the study initiated in \citet*[hereafter B05]{bro05}, to revisit the possibilities of extracting reliable information about past merger events in the halo of our Galaxy in the upcoming Gaia database.
In the previous study, we developed the numerical machinery to build `mock' Gaia catalogues that include merger events from $N$--body simulations against a realistic smooth Galactic background. Spatial, velocity and photometric information from a Galaxy model with three separate components (bulge, disc and halo) was used. Sampling issues were tackled, like the variation in probing depth of the satellite stellar luminosity function, as a function of position along a tidal streamer. A realistic model for Gaia errors was included too. With the resulting catalogue, a re--examination of the integral--space method of \citet{hel00} was done and it was concluded that, although promising, in real practice observational errors and the presence of an overwhelming Galatic background can compromise the efficiency of this method and great care in its use must be exercised. It is clear that we must use as many search techniques as possible, preferably those that include as much observationally available information as possible.

Following this path, we have decided to investigate now the `Great Circle Cell Count'  (GC3) method proposed by  \citet*{joh96}, in order to find a way to increase its efficiency. We propose an extension to this method, that we dubbed mGC3
(`modified GC3'), which adds kinematical information to the original method, in order  to detect substructures in the MW stellar halo. We have used the mock Gaia catalogue of B05, with the further improvement of a more detailed modeling of the simulated photometry of the satellites, so they now reflect specific histories of star formation. 
The expected Gaia measurement errors were simulated as a function of magnitude,
colour, and sky position of the stars, as described in appendix A of B05. The
values of the expected errors were updated to the latest predictions of the
scientific performance of Gaia. The latter can be found at
http://www.rssd.esa.int/gaia under the `Science performance' entry. Compared to
the performance predictions used in B05 the current astrometric error estimates
are larger by a factor of two.
We specifically asses the effect of the predicted Gaia observational errors on the ability of GC3/mGC3 strategies  to find simulated satellite remnants. We also study the effects on recovery efficiency when varying the orbit, luminosity and star formation history of the satellites.

In section~\ref{s:gc3}, we present a brief outlook on the problem of identifying past merger events and review the original GC3 method. Some specific examples are shown to illustrate how it works. In section~\ref{s:mgc3} the mGC3 extension is introduced. The basic concept, as well as a practical way to implement it, are discussed. In section~\ref{s:appl}, we evaluate the applicability of the mGC3 method using the mock catalogue of B05. We also describe the way in which we have included various star formation histories for the satellites using and adaptation of the stellar population synthesis software of \cite{bru03}.
In section~\ref{s:detectability}, we study in detail the efficiency of the mCG3 method in identifying past merger events, as a function of satellite orbit, luminosity and star formation history. Our conclusions are presented in section~\ref{s:conclusions}.

\section{The Great Circle Cell Count method}\label{s:gc3}
 
Current computer capabilities allow us to explore the formation and evolution of stellar streams under
a variety of conditions, such as different orbits and underlying potentials, via N-body simulations.
The understanding of common properties of stellar streams, as well as their time evolution, are key 
to develop effective identification techniques. This goes hand in hand with the
observational developments that will allow the practical application of such techniques to real data.

Key properties that determine the dynamical structure and evolution of  stellar streams, which in turn
can be used in their identification,  are the integrals of motion: the total energy $E$, if the Galatic potential
is stationary; the total angular momentum  $L$, if the potential has spherical symmetry;  and its projection
$L_z$, if the potential has axial symmetry.

\begin{figure*}
\includegraphics[width=180mm]{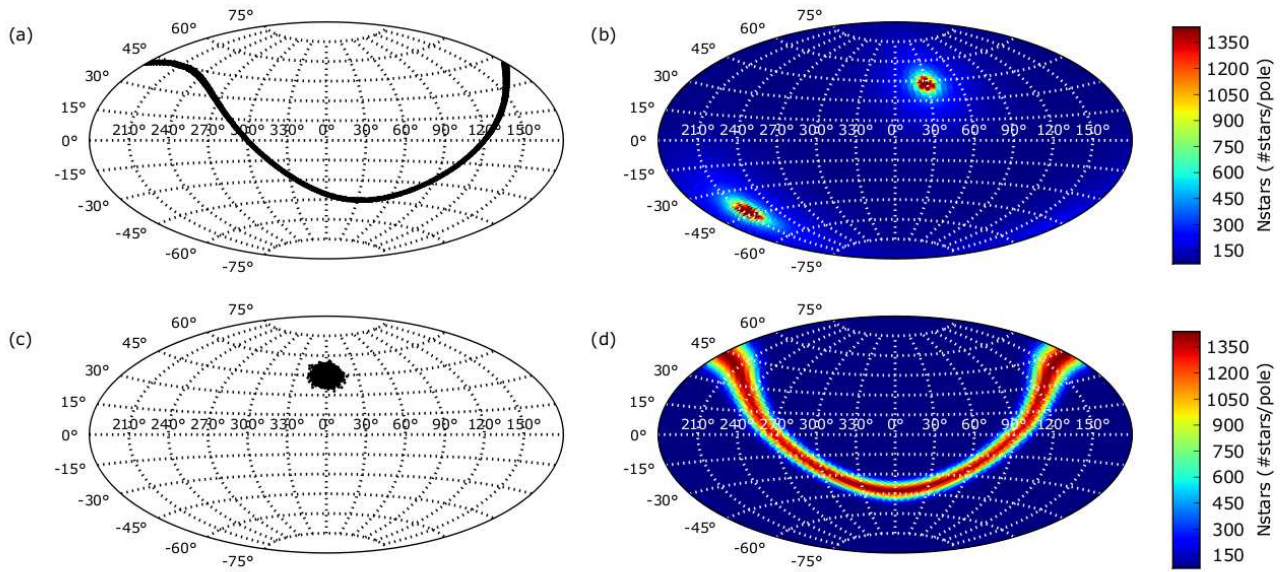} 
 \caption{Illustration of spatial and corresponding GC3 pole count maps of simple stellar distributions, shown in Aitoff projections. (\emph{a}) Spatial distribution of $9\times10^3$ simulated stars in a planar circular orbit, with an inclination of $45^\circ$ with respect to the Galactic Plane. (\emph{b}) GC3 pole count map of stars shown in (a). (\emph{c}) Spatial distribution of $9\times10^3$ simulated stars with a localized gaussian distribution. (\emph{d}) GC3 pole count map of stars shown in (d).}
 \label{fig:simple_gc3}
\end{figure*}

The existence of these conserved quantities leads naturally to the search of streams as entities in integrals
of motion space, as proposed by  \citet{hel00}. This technique requires the computation
of the conserved quatities in terms of observed ones, like $(l,b,\varpi, v_r,\mu_l,\mu_b)$, the Galactic latitude
and longitude, parallax, radial velocity and proper motions, respectively. These computations  involve non-linear 
transformations resulting also in non-linear  error propagation which, as B05 show, lead to systematic
errors in $E, L, L_z$ as their expectation values do not equal their true values once the observational errors
are introduced. Furthermore, the spread of errors is greatly magnified, as shown there.
Additionally, the computation of the total energy $E$ requires an assumption about the underlying potential,
which is an unknown and thus introduces another source of error. It is clear that, although powerful in theory, 
the integrals of motion technique has its limitations when applied to real data. It is thus convenient to look for
additional search techniques that may complement it.

A different approach in the identification of tidal streams is the GC3 method proposed by \citet{joh96}, 
which uses the conservation of the total angular momentum in a geometrical way.  It relies on the fact
that the orbits of stars in a stream in a spherical potential will be confined to a plane that contains  the
potential center.  As seen from the Galactic center, this plane is a great circle on the sky. In the more
general case of an axisymmetric Galactic potential, the orbital plane precesses around the axis of
symmetry of the system. This will result in the stream being approximately  confined to a plane with
a finite thickness, which in turn corresponds  to a great circle of finite width.  A further complication arises
from the fact that we are not at the Galactic center, which spoils to some degree the great circle effect,
particularly for streamers whose orbits lie within, or near the Sun's orbit.

The practical implementation of the GC3 method begins by laying down a grid on the celestial
sphere. Each grid point is then associated to a unit vector $\hat{L}$ (pole vector) that points from the
observer to the grid point. All stars on the sky that lie on a band defined by the great circle
in the plane orthogonal to the pole vector, are associated to this pole or grid point, the angular
width of the band being specified independently. The number of stars within this band are assigned to
the corresponding pole. The counts associated to every grid point in the mesh thus define a \emph{pole count
map}. Significant maxima in this pole count map are then accepted as possible streamers. Because of the
geometry, the pole vector will be parallel to the angular momentum of all stars in the band whose velocity
vector also lies in the orbital plane defined by the great circle. This latter property will be exploited later
on, when we introduce the extended GC3 method.

\begin{equation}\label{rcriterion}
|\mathbf{\hat{L}} \cdot \mathbf{\hat{r}}|  \leq \delta_r
\end{equation}

where $\mathbf{\hat{r}}$ is a unit vector pointing to the star and
$\delta_r=\sin\delta\theta$ is the tolerance which allows for
the width $\delta\theta$ of each great circle  associated to a cell \citep[see Sec. 3.1 and Fig. 7 in][]{joh96}. 
The distribution of pole counts is then represented in a map of the celestial sphere.

We illustrate the pole count maps for two simple \emph{simulated} stellar
distributions in Fig. \ref{fig:simple_gc3}a-d, in order to explain their morphology.
The spatial distribution of $9\times10^3$ stars in a circular orbit with a
$45^\circ$ inclination is shown in Fig. \ref{fig:simple_gc3}a 
and the corresponding pole count map in Fig. \ref{fig:simple_gc3}b. This shows how the
signature in a pole count map, of stars in a planar orbit, corresponds
to a localized peak which coincides with the direction of the
normal vector defining the orbital plane. Since two antiparallel pole vectors  
 define the same plane or great circle in the celestial sphere, 
the information on one hemisphere of a pole count map is replicated on the other;
hence the two peaks at exactly opposite directions in Fig.\ref{fig:simple_gc3}b.
Figure \ref{fig:simple_gc3}c shows a map of stars in a localized stellar
distribution, with the corresponding pole count map in  
Fig. \ref{fig:simple_gc3}d. The maxima in pole counts are in this
case distributed along a great circle. This happens because
all the possible planes which contain both, the localized distribution 
of stars and the observer at the Galactic Center, have poles which are in turn
contained in a specific plane, of which the projection on the sky is
the great circle seen in Fig. \ref{fig:simple_gc3}d. 

In the GC3 method it is expected that the stars in the Galactic background
will contribute a smooth distribution across the entire map, modulated only
by the stellar density dependence on position in the sky. The stars  in a stream,
on the other hand, will contribute only to the cells which coincide best  with the
stream's orbital plane, creating a local maximum in the pole count distribution. 

The GC3  method was used on a C-star survey
by \citet{iba01} in order to quantify the statistical significance
of a great circle C-star overdensity and to identify it
as part of the tidal stream from the Sagittarius (Sgr) dSph galaxy.
Later, it was used on 2MASS M-giants by \citet{iba02}, on a search for tidal debris
in the Galactic Halo. They succesfully recovered the Sgr
stream but failed to indentify any other significant features. 

In the following subsections we explore the performance of the original 
GC3 method by applying it to a simulated, or mock Gaia catalogue (B05).
The catalogue includes observational uncertainties and
a realistic number of stars in the Galactic background.
It illustrates the need to modify the original method in order
to reduce the contribution of the Galactic background,
and increase its sensitivity to detect lower luminosity 
tidal streams.

\subsection{The Mock Gaia Catalogue}\label{s:mock}

The Galactic catalogue from B05 constitutes 
a random realization of kinematic properties of 
Milky Way stars in the Galactic disc, bulge and halo, with 
their corresponding density,
velocity, age and color distributions, as well as
the appropriate normalization in total luminosity.
This resulted in a mock Gaia catalogue containing 
$\sim3.5\times10^8$ \emph{observable} Milky Way stars 
with full phase-space information and 
observational errors as are expected from Gaia, 
as well as the error-free quantities for comparison.
The two most important aspects to be emphasized 
about the mock catalogue are the realistic simulation
of both the observational errors and the total number 
of stars that will be observable with Gaia, which will
be key issues in determining the applicability of any
stream searching method.

\subsection{GC3 Galactic pole count maps}

We restricted the catalogue to stars with $|b|>10^\circ$,
in order to avoid the Galactic Plane. The tolerance used in 
(\ref{rcriterion}) was $\delta\theta=5^\circ$ which
corresponds to the half-width of each great circle cell and
the GC3 pole counts were computed on a 
$72 \times 72$ cell grid, uniformly spaced on the surface
of the celestial sphere. 
We chose the tolerance to be slightly less than the
$\delta\theta=6^\circ$ half-width which \citet{iba02}
find to maximize the signal to noise ratio
of the Sgr tidal stream feature in their 2MASS M-giant
pole count maps.

\begin{figure}
\includegraphics[width=84mm]{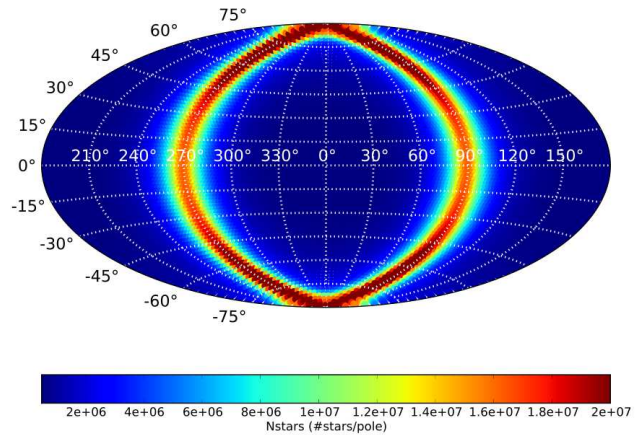}
 \caption{GC3 pole count map of Galactic background from error-free mock Gaia catalogue, in a Galactocentric reference frame. The color scale represents the number of stars per pole as shown in the color bar.}
 \label{fig:gc3_gal_true}
\end{figure}

We computed the pole counts for the error-free mock
Gaia catalogue, using the GC3 method's position criterion expressed in (\ref{rcriterion})
in a Galactocentric reference frame,
i.e. position vectors and pole coordinates are Galactocentric, the latter corresponding 
to normal vectors of planes that go through the Galactic center. The resulting
pole count map is shown in Fig. \ref{fig:gc3_gal_true}.
In this reference frame we name the longitude and latitude angles $\phi$ and $\theta$ respectively,
with the Galactic Plane  at $\theta=0^\circ$, the North Galactic Pole at $\theta=+90^\circ$
and $\phi=0^\circ$ in the direction away from the Sun.

\begin{figure}
\includegraphics[width=84mm]{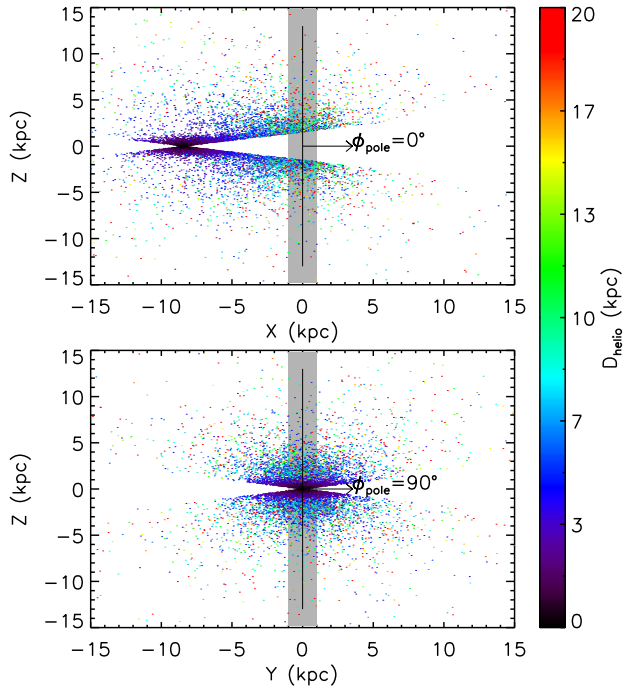} 
 \caption{Projections of mock Gaia catalogue stars. The color scale indicates heliocentric distances $D_\mathrm{helio}$ as indicated by the color bar.  \emph{Top:} X-Z projection. The grey shaded area shows the plane with pole  $\phi_{pole}= 0^\circ$ and $\theta=0^\circ$, which does not contain solar neighborhood stars. \emph{Bottom:} Y-Z projection. The grey shaded area shows the plane with pole  $\phi_{pole}=90^\circ$ and $\theta=0^\circ$. This plane contains solar neighborhood stars. Also, as shown by the color scale, the heliocentric distances of stars in the shaded area of the top panel are on average larger than those in the shaded area in the bottom panel. These effects give rise to the morphology noted on Fig. \ref{fig:gc3_gal_true} (see text for details).}
 \label{fig:gc3_planes}
\end{figure}

In this reference frame the pole count map would be expected to be
uniform in $\phi$ and with a positive gradient in $\theta$ (minimum at the equator),
because of the axial symmetry of the Galaxy about the Galactic center and the latitude dependence
in star density. However, the pole count map shown in Fig. \ref{fig:gc3_gal_true} clearly does not follow
this pattern in $\phi$.
This is due to both the exclusion criterion in Galactic latitude ($|b|>10^\circ$)
and the fact we are considering only stars which will be \emph{observable} by Gaia; 
both criteria being inherently heliocentric. 
First, the exclusion criterion imposed on the heliocentric 
Galactic latitude does not filter stars in the Galactic Plane near the Sun at high latitudes, 
 these stars contribute to pole counts for poles in the redish circle in Fig. \ref{fig:gc3_gal_true}, with their
maximum contribution being for  $\phi_{pole}\sim90^\circ$ and $\phi_{pole}\sim270^\circ$. 
In addition, planes with $\phi_{pole}=0^\circ,180^\circ$ 
do not contain solar neighbourghood stars and thus include stars at heliocentric 
distances that are larger on average than for planes that do go through the Sun with 
$\phi_{pole}=90^\circ,270^\circ$ (this is illustrated for $\phi_{pole}\sim0^\circ$ and $\phi_{pole}\sim90^\circ$ in Fig. \ref{fig:gc3_planes}). Therefore the number of observable
stars that contribute to poles $\phi_{pole}\sim90^\circ,270^\circ$ will be larger 
than in the perpendicular direction. These two effects reinforce
one another and give rise to the morphology of the
pole count map of Fig. \ref{fig:gc3_gal_true}.

The GC3 method's capability for detecting tidal streams clearly depends
on how large is the contribution of the stream with respect
to the Galactic background. In Fig. \ref{fig:gc3_gal_true}, the mode
of pole counts is $\sim4.5\times10^6$ stars/pole, which is a measure of 
the typical contribution of Galactic stars in this error-free map. 
On the other hand, a dwarf galaxy with a stellar population similar
to the Galactic Halo (i.e. an age of 13.6 Gyr and iron abundance $[\mathrm{Fe}/\mathrm{H}]=-1.7$)
and total luminosity of $10^7$ L$_\odot$ has $\sim1.2\times10^6$ stars
brighter than $M_V=5$, according to the stellar population synthesis
models of \citet{bru03}. The number of \emph{observable} 
stars, however, will depend on the distance distribution
of the stars in the stream, but will necesarily be smaller, specially since
most of these are rather faint ($M_V \gtrsim 4$) main-sequence stars. 
Therefore, the signature of such a dwarf galaxy in a pole count
map would be several times below the typical number of Galactic
background stars and thus very hard to detect.

As exhibited in this example, the number of  Galactic 
contaminants in a cell is an important limitation
of the GC3 method. In the following section we propose a 
modification of the method, to help discriminate 
stream stars from the smooth halo using the main attribute
of the original GC3 method, namely the use of
angular momentum conservation
through a geometrical approach.

\section{The Modified GC3 Method}\label{s:mgc3}

The GC3 method of Johnston et al., as explained above,
uses solely positional information. In order to improve
the `signal' of a stellar stream over the `noise'
of the smooth background halo population, we 
propose the inclusion of velocity information as well.
Since the stellar stream is roughly confined to a plane
that contains the center of symmetry of the potential,
as seen from this center both its position and velocity 
vectors will be contained in the orbital plane. 

A grid of all the possible great circle cells is devised,
with each cell uniquely determined by its \emph{pole}, 
that is the unit vector $\mathbf{\hat{L}}$ which is perpendicular to the plane it defines.
The modified GC3 (mGC3) method proposed here for counting the 
number of stars associated with each pole, can be expressed 
in terms of the following Galactocentric position and velocity criteria

\begin{equation}\label{criteria}
|\mathbf{\hat{L}} \cdot \mathbf{\hat{r}}| \leq \delta_r \quad \mathrm{and} \quad  |\mathbf{\hat{L}} \cdot \mathbf{\hat{v}}| \leq \delta_v 
\end{equation}

where $\mathbf{\hat{r}}$ and $\mathbf{\hat{v}}$ are unit vectors parallel to the star position and velocity vectors,
$\delta_r$ and $\delta_v$ are the tolerances which allow for
the width of each great circle associated with a cell. The pole vector $\mathbf{\hat{L}}$ that corresponds 
to the cell which best coincides with a stream is parallel to the
streams's total agular momentum.

As with the original GC3, the mGC3 criteria in (\ref{criteria}) will hold  for tidal
streams evolved in a spherical or slightly flattened potential.
For the Milky Way (MW), according to several authors, 
the orbit of the Sgr tidal stream constrains the inner halo ($\lesssim 60$kpc)
to be slightly flattened with a Z-axis flattening $q_z=0.85-0.95$  
for oblate models (\citealt*{joh05}; \citealt{md04,iba01}), and $q_z=1.25$ 
for prolate models \citep*{hel04,law05}; albeit more recently  \citet*{law09}
find that a triaxial halo with $q_z=1.25$ and $q_y=1.5$ provides a better fit of
the radial velocity and distance distribution of observed Sgr stream stars. 
On the other hand, depending on dynamical age, orbital inclination and how
close the stellar system comes to the Galactic plane, the 
disc's potential can perturb the symmetry of the overall 
potential felt by the stream. \citet{joh08} classify  
stream morphologies resulting from their N-body simulations 
in three categories `great circle', `cloudy' and `mixed', illustrated
in their Fig. 1, as well as transition types in between these
categories, as shown in their Fig. 2. The mGC3 criteria proposed here will hold for the 
`great circle' streams, as well as for the dynamically younger parts 
of `cloudy' and `cloudy-great-circle' streams, and clearly not for 
`mixed' morphology streams.

\subsection{Practical Implementation of the mGC3 criteria}\label{s:mgc3_practical_impl}

The position and velocity criteria proposed above, need to be written in terms
of observable quantities. Our first step is to write them using the Galactocentric
position and velocity vectors of each star:

\begin{equation}\label{unit_criteria}
|\mathbf{\hat{L}} \cdot \mathbf{r}_\mathrm{gal}| \leq \Vert \mathbf{r}_\mathrm{gal} \Vert\delta_r \quad \mathrm{and} \quad  |\mathbf{\hat{L}} \cdot \mathbf{v}_\mathrm{gal}| \leq  \Vert \mathbf{v}_\mathrm{gal} \Vert \delta_v \,,
\end{equation}

where $\delta_r=\sin\delta\psi_r$, $\delta_v=\sin\delta\psi_v$ and $\delta\psi_r, \delta\psi_v$ 
are the complements of the angles between $\mathbf{\hat{L}}$ and the vectors $\mathbf{r}_\mathrm{gal}$, and  
$\mathbf{v}_\mathrm{gal}$ respectively, which correspond to the tolerance width of the great 
circle associated with the cell.

The vectors  $\mathbf{r}_\mathrm{gal}$ and $\mathbf{v}_\mathrm{gal}$ are in turn expressed in
terms of the observable quantities $(l,b,\varpi, v_r,\mu_l,\mu_b)$ as follows

\begin{equation}\label{rgal_vgal}
 \begin{array}{l}
  \mathbf{r}_\mathrm{gal} = \mathbf{r}_\odot + A_p\varpi^{-1}\big[ (\cos{l}\cos{b}) \mathbf{\hat{x}}+(\sin{l}\cos{b}) \mathbf{\hat{y}}+(\sin{b}) \mathbf{\hat{z}} \big ] \\
 \mathbf{v}_\mathrm{gal} =  \mathbf{v}_\odot + v_r \mathbf{\hat{r}} + \varpi^{-1} \big [ (A_v \mu_l \cos{b}) \mathbf{\hat{l}}+ (A_v\mu_b) \mathbf{\hat{b}} \big] \,,
 \end{array} 
\end{equation}

where $A_p=10^3$ mas$\cdot$pc, $A_v=4.74047$ yr \kms; $\lbrace \mathbf{\hat{x}},\mathbf{\hat{y}},\mathbf{\hat{z}} \rbrace$ are the unit vectors in the cartesian Galactocentric reference frame and $\lbrace \mathbf{\hat{r}},\mathbf{\hat{l}},\mathbf{\hat{b}} \rbrace$ are the unit vectors in a spherical heliocentric reference frame. The latter depend upon $(l,b)$, though not on the parallax $\varpi$, when expresed in the cartesian Galactocentric reference frame (see Appendix \ref{s:appendix}). 

As can be seen from (\ref{rgal_vgal}), the expressions for the Galactocentric position
and velocity vectors $\mathbf{r}_\mathrm{gal}$ and $\mathbf{v}_\mathrm{gal}$ depend upon the reciprocal
of the parallax, which as discussed above, will introduce systematic errors in the determination
of both vectors. Therefore we define the vectors $\mathbf{r'}_\mathrm{gal}\equiv\varpi \mathbf{r}_\mathrm{gal}$, 
and $\mathbf{v'}_\mathrm{gal} \equiv \varpi \mathbf{v}_\mathrm{gal}$. With these definitions the parallax will
enter multiplying instead of dividing (see \ref{rgal_vgal_prime}).

\begin{equation}\label{rgal_vgal_prime}
\begin{array}{l}
 \mathbf{r'}_\mathrm{gal} = \varpi\mathbf{r}_\odot + A_p(\cos{l}\cos{b}) \mathbf{\hat{x}}+(\sin{l}\cos{b}) \mathbf{\hat{y}}+(\sin{b}) \mathbf{\hat{z}} \\
\mathbf{v'}_\mathrm{gal} =  \varpi\mathbf{v}_\odot + \varpi v_r \mathbf{\hat{r}} + (A_v \mu_l \cos{b}) \mathbf{\hat{l}}+ (A_v\mu_b) \mathbf{\hat{b}}
\end{array} 
\end{equation}

We can thus rewrite the criteria of (\ref{unit_criteria}) in terms of the 
vectors $\mathbf{r'}_\mathrm{gal}$ and $\mathbf{v'}_\mathrm{gal}$ as shown in  (\ref{mod_criteria}),
which will constitute our working criteria.

\begin{equation}\label{mod_criteria}
|\mathbf{\hat{L}} \cdot \mathbf{r'}_\mathrm{gal}| \leq \Vert \mathbf{r'}_\mathrm{gal} \Vert\delta_r \quad \mathrm{and} \quad  |\mathbf{\hat{L}} \cdot \mathbf{v'}_\mathrm{gal}| \leq  \Vert \mathbf{v'}_\mathrm{gal} \Vert \delta_v 
\end{equation}

These expressions are numerically equivalent to those in (\ref{unit_criteria}), 
though they are in practice more useful, since the effect of error propagation due to the reciprocal of the parallax is avoided. The method based on these criteria relies on an accurate knowledge of $\mathbf{r}_{\odot}$ and
$\mathbf{v}_{\odot}$, which we can expect to be accurately determined from Gaia data.

\subsection{Modified GC3 applied to the mock Gaia Catalogue}

We computed the pole counts for the mock Gaia
catalogue described in Sec. \ref{s:mock}, now using
the modified GC3 (mGC3) criteria expressed in (\ref{mod_criteria}). 
As in Sec. \ref{s:gc3}, we restricted the catalogue to stars 
with $|b|>10^\circ$ and the mGC3 pole counts were computed on a 
$72 \times 72$ cell grid, with equal tolerances for the
position and velocity criteria $\delta\psi_r=\delta\psi_v=5^\circ$.

\subsubsection{mGC3 pole counts of mock catalogue without errors}\label{s:mgc3_galmap_noerrs}

\begin{figure}
\includegraphics[width=84mm]{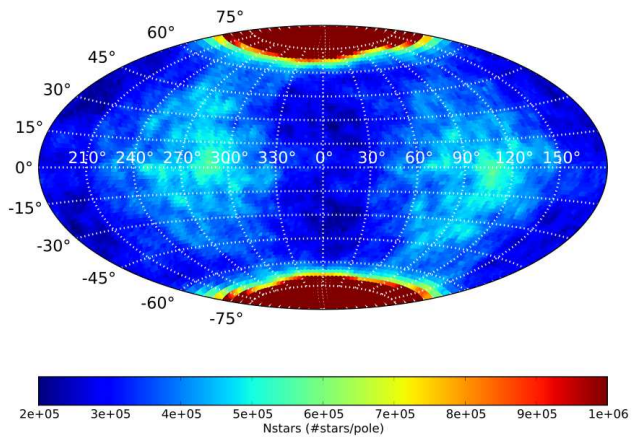}
 \caption{Map of mGC3 pole counts for mock Gaia catalogue \emph{without} observational errors. The color scale represents the number of stars per pole as shown in the color bar. Notice the change in normalization with respect to Fig. \ref{fig:gc3_gal_true}}
\label{fig:mgc3_gal_true}
\end{figure}

For the error-free mock catalogue the resulting pole count map is shown
in Fig. \ref{fig:mgc3_gal_true}. In this map the mode of pole counts is $\sim 4\times10^5$,
a factor of $10$ lower than the typical pole counts in the GC3 map
of Fig. \ref{fig:gc3_gal_true}.This is a consequence of the velocity 
requirement of (\ref{mod_criteria}), which decreases the
overall number of stars per pole because the contribution to any 
given pole, or great circle cell, does
not come from all the stars in the plane of the great circle associated with the cell, but comes
only from those whose velocity is also contained in the great circle plane,
therefore decreasing the contribution to each 
pole by chance alignments. This effectively lowers the pole counts
on the whole map and also reduces the strength of the pattern of
excess pole counts around $\phi_{pole}=90^\circ,270^\circ$ because,
as discussed in Sec. \ref{s:gc3}, the majority of stars that contribute
to this excess are in the Galactic Plane,
and so their velocities do not lie on the plane associated to these $\phi_{pole}$ values.

\subsubsection{mGC3 pole counts of mock catalogue with errors}\label{s:mgc3_galmap_errs}

\begin{figure}
\includegraphics[width=84mm]{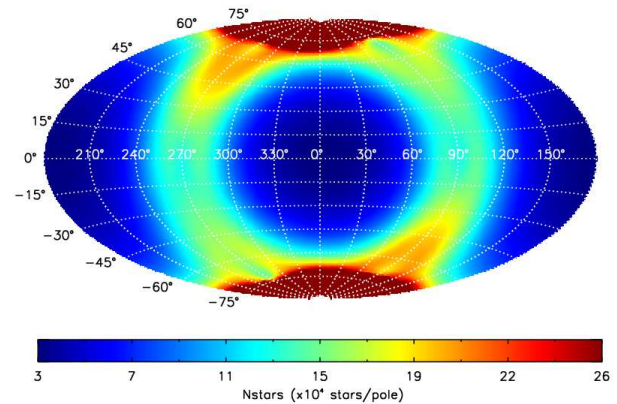} 
 \caption{Map of mGC3 pole counts for mock Gaia catalogue \emph{including} simulated observational errors.The color scale represents the number of stars per pole as shown in the color bar. Note that the maximum of the color scale in this figure is $\sim10$ per cent of Fig. \ref{fig:mgc3_gal_true}.}
\label{fig:mgc3_gal_obs}
\end{figure}

The pole count map for the mock catalogue \emph{including}
the simulated errors is shown in Fig. \ref{fig:mgc3_gal_obs} and
was computed using only stars with $|b|>10^\circ$ and with positive 
parallax with errors smaller than $30$ per cent. 

This requirement on the parallax causes a pattern of excess
pole counts for poles at $\phi_{pole}\sim90^\circ,270^\circ$,
similar to that seen in the GC3 pole count map discussed 
in Sec. \ref{s:gc3}. The origin of this pattern is essentially
the same as discused in Sec. \ref{s:gc3}, great circle cells
with poles $\phi_{pole}\sim0^\circ,180^\circ$ include 
stars with heliocentric distances that are on average larger than
for perpendicular cells; since parallax errors increase
with heliocentric distance, this causes less stars to pass the 
parallax error filter for cells that do not include the Sun, 
and conversely an excess for cells
that do ($\phi_{pole}\sim90^\circ,270^\circ$). 
However, the background level is vastly reduced compared to that of Fig. \ref{fig:gc3_gal_true}.

\section{Applicability of the mGC3 method}\label{s:appl}

We now evaluate the applicability of the mGC3 method by applying it to pole count maps
including the Galactic background from the mock Gaia catalogue described in Sec. \ref{s:mock}
and tidal streams from N-body simulations also from B05. In the
following section we describe the kinematic and photometric characteristics
used in our experiments and the morphology of mGC3
pole count maps is illustrated in detail in Sec. \ref{s:typical_stream} for a typical tidal stream.

\subsection{Assigning observable properties to simulated satellite streams}\label{s:obsprops}

We use the simulated satellites from B05 (see paper for details), which are the results of
N-body simulations of self-consistent systems with $10^6$ particles evolved in a rigid axisymmetric
Galactic potential. The Galactic potential used was oblate with a
flattening in the potential of $q_{\Phi_h}=0.8$. 
The satellites have two different initial masses of $2.8\times10^7$ M$_\odot$
and $5.6\times10^7$ M$_\odot$ and different pericentre, apocentre and inclination angles with respect to the Galatic plane, which are summarized in Table \ref{tab:orb_params}. The satellite IDs used here (S1-S5)
correspond to the Run No. (1-5) used in Table 6 of B05.
The resulting orbits are illustrated in Fig. 4 in B05. For each satellite, snapshots of the
distribution of N-body particles are available for dynamical ages ranging from 0 to 9.75 Gyr 
at 0.65 Gyr steps. We use the term \emph{dynamical age} to emphasize this 
does not refer to the age of the stellar population, but to the elapsed time in the
N-body simulation. We take the dynamical age to be zero in the beginning of the
N-body simulation, when the satellite is entirely bound.

\setcounter{table}{0}
\begin{table}
\caption{Orbital parameters of simulated satellites}
\begin{tabular}{ccccc}
\hline
Sat & Mass                   & Pericentre & Apocentre & Inclination \\
ID   & ($\times10^7$ M$_\odot$)          & (kpc)          & (kpc)         & angle ($^\circ$) \\ 
\hline
S1    & 5.6 & 8.75           & 105          & 30 \\
S2    & 5.6 & 7.0             & 60            & 45 \\
S3    & 2.8 & 7.0             & 80            & 60 \\
S4    & 2.8 & 40              & 60            & 25 \\
S5    & 2.8 & 3.5             & 55            & 45 \\
\hline
\end{tabular}
\label{tab:orb_params}
\end{table}

Embedding the simulated satellite streams in the Galactic
background described in Sec. \ref{s:mock}, requires that we make a realistic simulation 
of the number of stars that will be observable by Gaia in each of the simulated satellites. 
This number depends on the total luminosity $L_V$ of the satellite;
on its orbit, which determines the distribution
of observer to star distances along the stream; and on
its star formation history (SFH), which will determine the luminosity
function of the stellar population at any given age.

For our experiments we varied the total luminosity $L_V$ 
in the range $1\times10^7$ L$_\odot$ to $1\times10^9$ L$_\odot$.
In all cases the luminosity $L_V$ refers to the total luminosity
of the stars in the whole satellite system including the tidal tails,
not just the bound core.

For each simulated satellite we assigned two different SFHs, 
whose properties are described in Table \ref{tab:sfhs}. 
These are termed Halo-type and Carina-type since they
have been tailored to resemble the SFHs of the Galactic Halo and Carina dSph satellite, respectively. 
As can be seen from Table \ref{tab:sfhs}, the Halo-type SFH consists of only
an old burst with an age of 13 Gyr, with [Fe/H] near the peak 
of the iron abundance distribution of halo stars ([Fe/H]=-1.8, \citealt{Prantzos09}). 
The Carina-type SFH consists of three separate
bursts producing old (13 Gyr), intermediate-age (8 Gyr) and young (3 Gyr) populations, 
the intermediate and old being the dominant ones, and a single metallicity (although 
Carina exhibits a range in [Fe/H] from -2.0 to -1.0, see e.g. \citealt{Carigi2008,Carigi2002}). 
This choice of SFHs and [Fe/H] does not intend to accurately represent the stellar populations
of neither the present-day dSphs nor of the possible halo building-blocks. It only intends
to illustrate the performance of the mGC3 method in the least favorable case (the Halo-type SFH) 
and a slightly more favorable one (the Carina-type SFH), in terms of the number of 
stars observable by Gaia.

Using an adaptation of the \citet{bru03} stellar population synthesis software,
we generated a random realization of the photometric properties of stars in a system 
with total luminosity $L_V$ and a given SFH and age. This is done by selecting the SFH, 
and metallicity $Z$ (a metallicity Z=0.0004 corresponds to an iron abundance [Fe/H]=-1.7 at a solar [$\alpha$/Fe]  ratio). The ages of the stellar populations were fixed at the values given
by the SFHs in Table \ref{tab:sfhs}, independently of the dynamical age, for all experiments.
Luminosities and temperatures are then randomly drawn for as many stars as necessary, until 
the total luminosity reaches the desired value of $L_V$. Since a large number of the stars
in the realization will be low-mass stars too faint to be observed, we keep only 
the photometric information of stars brigther than an arbitrary cutoff at absolute magnitude $M_V=5$.
Although we discard these very faint stars, it is important to emphasize that their luminosities \emph{do}
contribute to the total luminosity of the system. 
The magnitude cutoff was chosen at $M_V=5$ since it is fainter than the turn-off of a 
13.4 Gyr old population of halo-like metallicity ($M_V=4$). This ensures that the turn-off 
will be brigther than our cutoff for a population of any given age. Nevertheless, the fraction of these
stars that will be observable will be determined by the Gaia magnitude cut-off ($V<20$).
Following this recipe ensures that our random realization accurately
represents both the luminosity function of the stars given by the SFH, as well as
the total number of stars in a system with total luminosity $L_V$. 

Finally, we randomly assign the photometric properties generated with this procedure, 
to the N-body particles of the simulated satellites as explained in Sec. 4.3 in B05, and
then compute, using the recipe detailed in Sec. 4.1 in B05, the Galatic longitude, 
latitude, parallax, radial velocities and proper
motions with their corresponding errors, for each simulated satellite star 
within the completeness limit of Gaia ($V<20$).

\begin{table}
 \begin{minipage}{100mm}
\caption{Star Formation Histories}
\begin{tabular}{lccc}
\hline
Name          & Age\footnote{Age of the star formation bursts assigned to simulated dwarf galaxies.}  & [Fe/H] & $L_V$ \\
                   & (Gyr) & (dex)  & (per cent) \\
\hline
Halo-type    & 13 & -1.7 & 100 \\
\hline
                   & 13 & -1.7 & 30 \\
Carina-type & 8   & -1.7 & 50 \\
                   & 3   & -1.7 & 20 \\
\hline
\end{tabular}
\label{tab:sfhs}
\end{minipage}
\end{table}

\subsection{Aplication to a typical stream}\label{s:typical_stream}

In this section we use a typical satellite stream with a chosen dynamical age,
luminosity, orbit and SFH in order to explain the morphology of the pole count map
and the methodology devised to detect in it the signature produced by the stream.
For the present and all following experiments, pole count maps were produced 
with a $72 \times 72$ spherically uniform grid, position and velocity tolerances
of $\delta\psi_r=\delta\psi_v=5^\circ$ and including from the simulated catalogues only those
stars with positive parallax with errors less than $30$ per cent and with $|b| > 10^\circ$, 
to avoid the Galactic Plane.

\begin{figure}
\includegraphics[width=84mm]{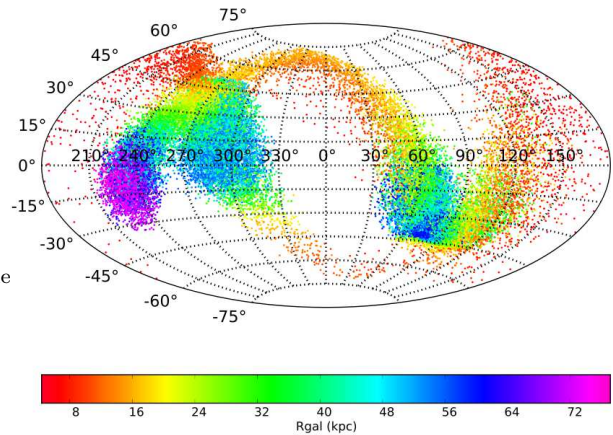} 
 \caption{Galactocentric sky distribution of N-body particles in satellite S2 at a dynamical age of 5.85 Gyr.The color scale represents Galactocentric distance as shown in the color bar.}
 \label{fig:spatial_s2}
\end{figure}

\begin{figure*}
\includegraphics[width=84mm]{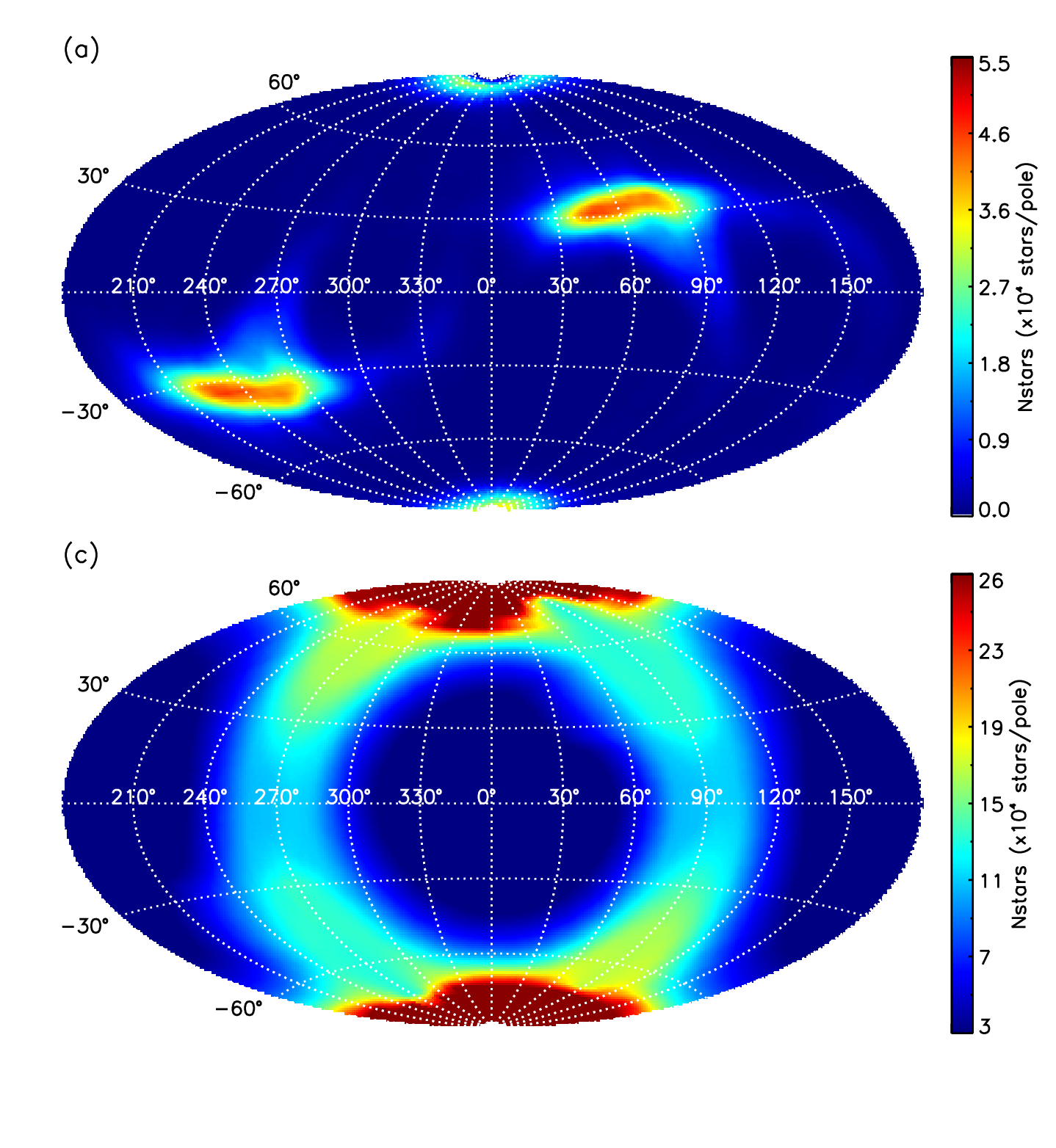}
\includegraphics[width=84mm]{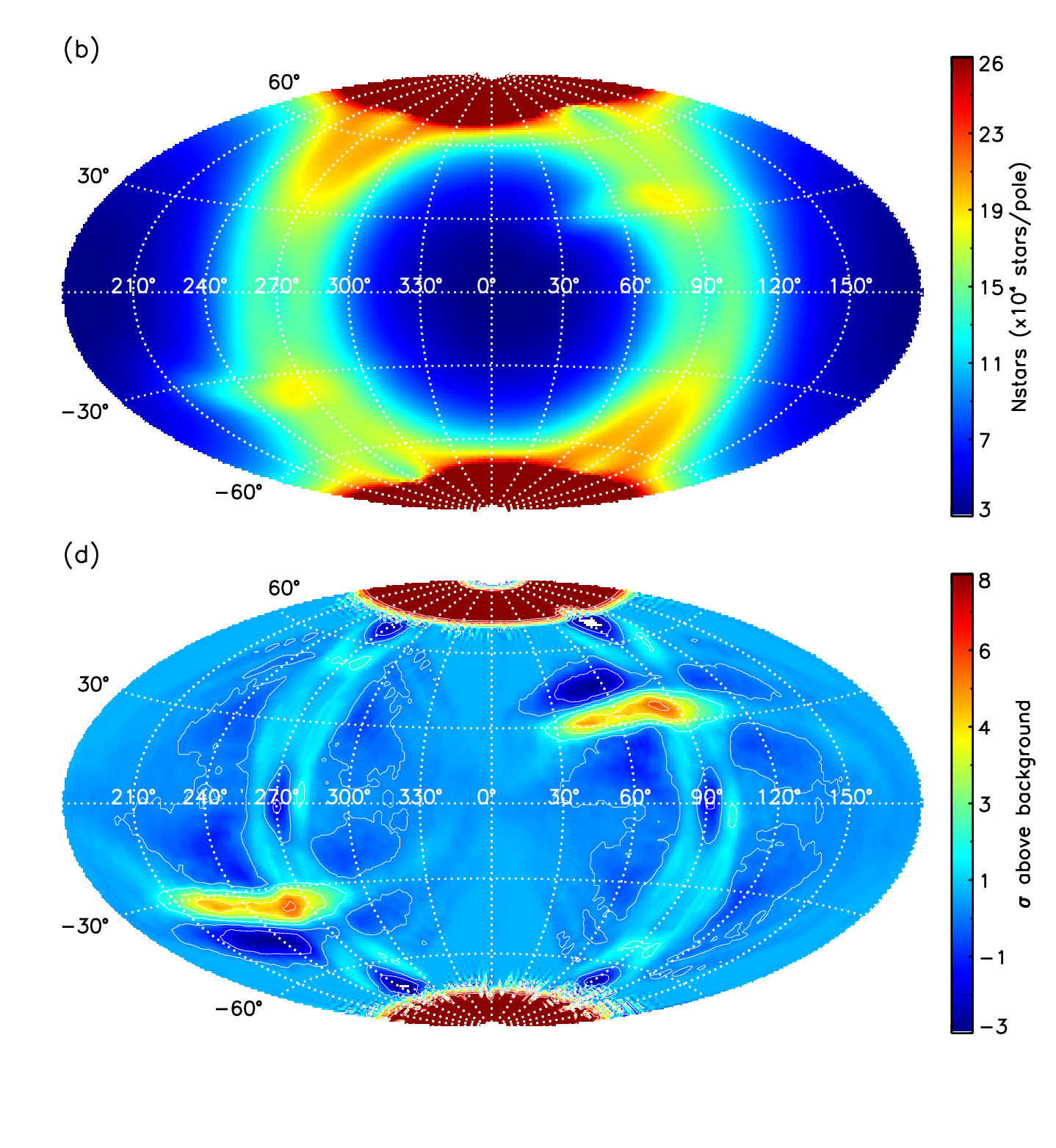} 
\caption{\textit{(a)} Map of mGC3 pole counts for stars in satellite S2 alone, at a dynamical age of 5.85 Gyr. \textit{(b)} Map of mGC3 pole counts for satellite S2 (panel a) embedded in the mock Galactic background. \textit{(c)} Smoothed image of the pole count map in panel (b) used for the `unsharp masking'. The color scales of panels (a), (b) and (c) represent the number of stars per pole as shown in the color bar. Note that the range of the color scale of panel (a) is 5 times smaller than that of panels (b) and (c). 
\textit{(d)} Subtracted pole count map. The color
scale and contours represent the amplitude or height of the excesses in units of the background's standard deviation $\sigma$.}
 \label{fig:polemap_s2}
\end{figure*}

For this example we chose satellite S2 at a dynamical age of 5.85 Gyr, 
with an elongated and inclined orbit, a Carina-type SFH and total luminosity of $3\times10^8$ L$_\odot$. The corresponding 
Galactocentric sky distribution is shown in Fig. \ref{fig:spatial_s2}. The color scale of the figure
indicates the particles' Galactocentric distances. The mGC3 pole count map which
corresponds to this simulated stream alone is shown in Fig. \ref{fig:polemap_s2}a; and the mGC3 pole count map of this stream embedded in the Galactic background (with errors) described
in Sec. \ref{s:mgc3_galmap_errs}, is shown in Fig. \ref{fig:polemap_s2}b. In this figure, the signature
of the stream in pole counts is a barely perceptible excess around 
$(\phi_{pole},\theta_{pole})=(85^\circ,35^\circ)$\footnote{The slight differences in the shape of map features with $\theta<0^\circ$ compared to those with $\theta>0^\circ$ are an effect of the Aitoff projection.}. 
This is the kind of feature we need to detect in pole count maps
in an automated fashion, in order to evaluate the method's capabilities 
to detect tidal streams of various characteristics. 

Since the  Galactic background leaves a smooth signature in the pole
count maps, we can use the standard image processing technique 
of \emph{unsharp masking} to remove its contribution. 
Unsharp masking consists in subtracting from the original image 
a \emph{smoothed} image, in which the value of each pixel corresponds 
to the median value in its neighbourghood. The subtracted image or,
in this case, the subtracted pole count map has a much more uniform background
and the contrast of localized excesses is enhanced. 
The smoothed image, made from the pole count map on Fig. \ref{fig:polemap_s2}b  
using a neighbourhood of $5\times5$ pixels, is shown in Fig. \ref{fig:polemap_s2}c.
The subtracted pole count map is shown in Fig. \ref{fig:polemap_s2}d for the present
example. In this subtracted map the stream's signature is much more evident
than it was in the original pole count map (Fig. \ref{fig:polemap_s2}b). The
color scale in the subtracted map indicates the amplitude or height of the
excesses in units of the background's standard deviation $\sigma$, and clearly
shows that the stream in this example is detected at the $\sim7\sigma$ level
in the subtracted pole count map; whereas in original map the excess appears only at
the $\sim2\sigma$ level. 

The subtracted pole count map of Fig. \ref{fig:polemap_s2}d still shows signs of
non-uniformity. Some regions in the subtracted map, particularly near an excess, appear
with pole counts below the background (i.e. negative sigma amplitudes 
in Fig. \ref{fig:polemap_s2}d). This is caused by an oversubtraction of the background
due to the excess itself, which increases the median value of the pole counts in 
pixels surrounding it in the smoothed image. Also, spurious detections tend
to arise at the $2-3.5\sigma$ level due to imperfections in the background subtraction.
In the examples of following sections we consider \emph{bona fide} excess detections
as those corresponding to excess counts with amplitudes larger than $4\sigma$. We also
require $|\theta_{pole}|<80^\circ$, consistently with our avoidance zone of $|b|>10^\circ$.

In the following section we systematically apply this procedure to recover the simulated streams from
the detection of excesses in mGC3 pole count maps.

\section{Variables affecting the detectability of tidal streams}\label{s:detectability}

In this section we explore the efficiency of the method in recovering
simulated streams under different conditions of total luminosity, SFH
and orbital parameters.

\subsection{Total Luminosity and Star Formation History}\label{s:det_LSFH}

\begin{figure}
\includegraphics[width=84mm]{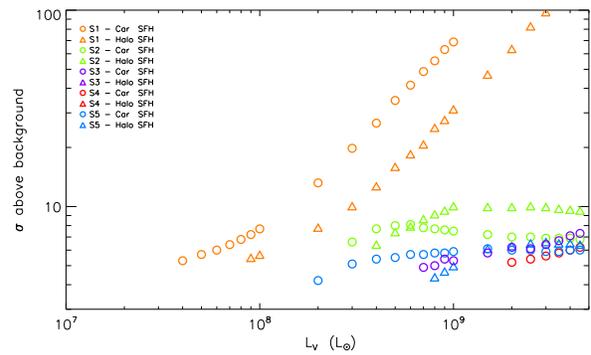} 
 \caption{Height of detected maxima on pole count maps as a function of the satellite's initial total luminosity $L_V$.}
 \label{fig:sigma_lum}
\end{figure}

The effect of increasing the total luminosity $L_V$ of a system is to increase the number
of observable stars proportionally. The stream's signature increases with
total luminosity, as can be seen in Fig. \ref{fig:sigma_lum}. 
For all five satellites, at an arbitrarily fixed dynamical age of 7.15 Gyr, Fig. \ref{fig:sigma_lum} 
shows the amplitude of the excesses recovered as a function of $L_V$.
The signal to noise ratio (SNR, understood as height of the peak over mean background,
in terms of the background's $\sigma$)  of the stream increases with $L_V$.
For some satellites, i.e. S2, the SNR reaches a plateau since, depending
on the shape of the stream, the pole count signature can differ from
a localized peak if the stream is very disrupted (see stream orbits in Fig.4 of B05). 
Therefore increasing the luminosity
can also increase the dispersion of pole counts in the background, 
maintaining the height approximately constant when expressed in terms
of the background's standard deviation, as in Fig. \ref{fig:sigma_lum}.

Figure \ref{fig:sigma_lum} also shows the effect of varying
the SFH, which in terms of the pole count maps is only
reflected in a change in the fraction of observable stars. At the same
total luminosity, a SFH having young or intermediate-age bursts (i.e. a Carina-like SFH)
yields a larger fraction of brigther stars compared to a 
population with only an old burst (i.e. an Halo-like SFH). 
This can be seen in Fig. \ref{fig:sigma_lum}, were for most satellites at a given luminosity
a stream is detected at a lower sigma level if the population is given
by the Halo-like SFH (triangles) than if it is given by a Carina-like SFH (circles).
For very disrupted satellites like S2 and S5, as mentioned before, 
the signature in pole count maps differs from a localized peak; therefore the significance
of the detection saturates earlier in luminosity for brighther populations causing
the stream to be detected at higher sigma levels for fainter populations.

Finally, it can be seen that satellite S1 can be detected with the mGC3 method
down to a luminosity as low as $\sim4\times10^7$ L$_\odot$, even
for this old dynamical age of 7.15 Gyr.

\subsection{Orbital parameters and dynamical age}\label{s:det_orbparams}

Orbital parameters are clearly important since they determine the shape of the orbit.
How disrupted the stream will become strongly depends on the perigalacticon and 
inclination angle. Nevertheless, the most influential parameter affecting the detectability
is rather the distribution of heliocentric distances of the stream stars at different dynamical 
ages, since this will determine the fraction of stars that will be observable in our simulated survey. 

\begin{figure}
\includegraphics[width=84mm]{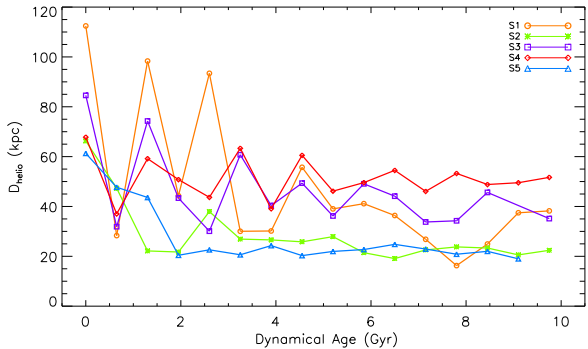} 
 \caption{Heliocentric distance $D_\mathrm{helio}$ as a function of dynamical age for all five satellites.}
\label{fig:perihel_age}
\end{figure}

For all dynamical ages of the five satellite's streams, we computed $D_\mathrm{helio}$
the mean of the heliocentric distance distribution, as a measure of a representative
heliocentric distance of the system. The plot of $D_\mathrm{helio}$ as a function
of dynamical age in Fig. \ref{fig:perihel_age}, shows that as the dynamical age
increases, stream stars can come closer to the Sun as they are
gradually distributed along the orbit, hence decreasing the mean of the distribution
of $D_\mathrm{helio}$. This effect would be enhanced by the action of dynamical friction
which, although not included in the simulations, is expected to 
contribute in the evolution of real systems.

\begin{figure} 
\includegraphics[width=84mm]{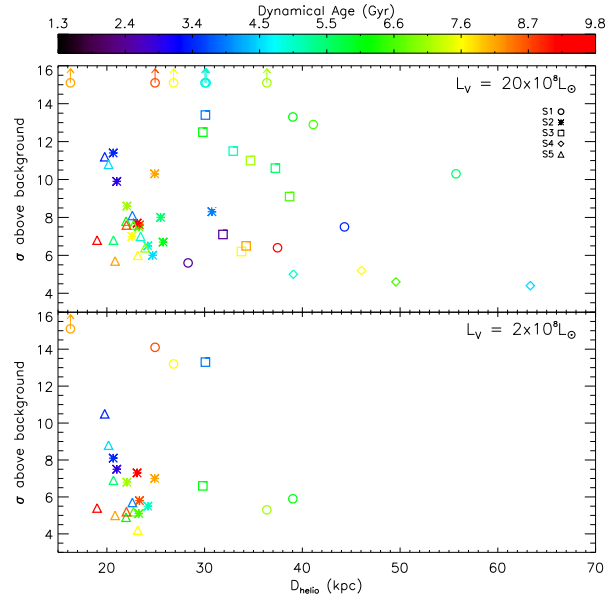}
 \caption{Height of detected excesses on subtracted pole count maps as a function of $D_\mathrm{helio}$.  Satellites have a Carina-type SFH and total luminosities $L_V=2\times10^8$ L$_\odot$ (\emph{Bottom}) and $L_V=2\times10^9$ L$_\odot$ (\emph{Top}). Detections with $\sigma>15$ are indicated with upward pointing arrows. The color scale represents dynamical age as shown by the color bar.}
\label{fig:sigma_perihel}
\end{figure}

Therefore, the detectability of streams should increase with decreasing $D_\mathrm{helio}$.
This is illustrated in Fig. \ref{fig:sigma_perihel}, a plot of the $\sigma$
detection level of the simulated satellite streams in pole count maps, 
as a function of $D_\mathrm{helio}$, for satellites with a Carina-type SFH and
two different luminosities $L_V=2\times10^8L_\odot$ (bottom) and $L_V=2\times10^9L_\odot$ (top).
These plots show that old streams ($\gtrsim7$ Gyr) 
at heliocentric distances up to $D_\mathrm{helio}\sim30$ kpc and $D_\mathrm{helio}\sim40$ kpc can be detected at a high sigma level, for the bright ($L_V=2\times10^9L_\odot$) and faint satellites ($L_V=2\times10^8L_\odot$) respectively.
On the other hand, streams can be  detected at larger distances only
if there's a dynamically younger and/or brighter population.
Also, at a given distance, streams with different dynamical ages (color scale) 
can be detected at different sigma levels.

\subsection{Combined Effects}\label{s:det_combined}

\begin{figure} 
\includegraphics[width=84mm]{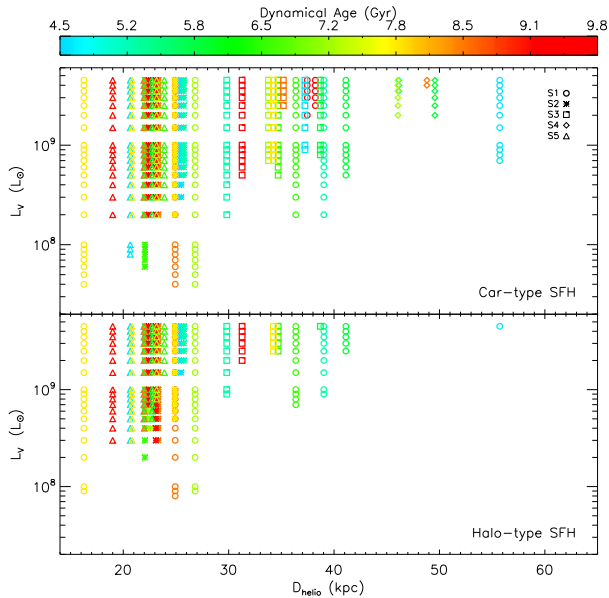}
 \caption{Total luminosity $L_V$ of detectable satellites as a function of the heliocentric distance, for dynamical ages greater than 4.5 Gyr, assuming a Halo-type SFH (\emph{bottom}) and a Carina-type SFH (\emph{top}). The color scale represents dynamical age as shown by the color bar.}
\label{fig:lv_perihel}
\end{figure}

In order to combine the different effects that impact the detectability
of simulated streams, explored individually in the previous sections, we show
in Fig. \ref{fig:lv_perihel} a plot of the total luminosity $L_V$ of detectable satellites 
as a function of $D_\mathrm{helio}$, for all five simulated satellites at dynamical 
ages larger than 4.5 Gyr and with Halo-type (\emph{bottom}) and Carina-type (\emph{top}) SFHs. 
This plot shows how all five satellites (S1-S5), 
which have stars observable by Gaia (and meeting our criteria $\sigma\varpi/\varpi\leqslant30$ per cent),
can be recovered using mGC3 at different dynamical ages, for total luminosities 
in the range $10^8-10^9$ L$_\odot$ with both SFHs and even down to $4-5\times10^7$ L$_\odot$ for satellites including a younger stellar population (Carina-type SFH) for certain combinations of dynamical ages and orbital parameters.

A statistical description of the efficiency of the mGC3 method would require
a more thorough exploration of the orbital parameter space, but in the present
experiments we have only 5 different orbital parameter sets. Nevertheless, these experiments
are usefull for a first rough exploration of the applicability of the mGC3 method in
which, as shown in Secs. \ref{s:det_LSFH} and \ref{s:det_orbparams}, 
we find that satellites with total luminosities
as faint as $\sim5\times10^7$ L$_\odot$, or as dynamically old as $\sim9$ Gyr,
with different orbital parameters, can be recovered. Furthermore, the experiments 
carried out illustrate a lower limit on the capabilities of the mGC3 method, since
in our simulations we use \emph{only} the information from trigonometric parallaxes
to obtain the distances to stars. Gaia, however, will also provide \emph{photometric} parallaxes
for a much larger, fainter and more distant sample of stars, than those for
which trigonometric parallaxes can be measured accurately.

\section{Discussion and Conclusions}\label{s:conclusions}

Motivated by the desire to unravel the fossil record of the formation of our
Galaxy we have developed an extension of the GC3 method of \cite{joh96} in order
to apply this to the search for satellite remnants in the Galactic halo. The GC3
method is essentially a means of grouping together stars on the sky with the
same orbital angular momentum, thus identifying stars along the same Great
Circle. The original method suffers from a strong contaminating background in
the pole count maps as demonstrated in Section \ref{s:mgc3}. These
background counts are due to the general Galactic population of stars and can be
much reduced by adding a kinematic criterion to the search for stars with similar
orbital angular momenta. This constitutes the extended mGC3 method presented in
this paper.

We applied this new method to a mock Gaia catalogue containing some
$3.5\times10^8$ stars that form the smooth Galactic background and stars from
simulated satellite galaxies on different orbits and with two types of star
formation histories. The results show that the mGC3 method is capable of tracing
remnants of satellites with luminosities down to $L_V\sim4-5\times10^7$ L$_\odot$
for dynamical ages up to $\sim 7$ Gyr. Remnants of brighter satellites
($10^8$--$10^9$ L$_\odot$) can be recovered up to dynamical ages of $\sim 10$
Gyr. The method works well for most satellites out to a heliocentric distance of
$40$ kpc. At larger distances only the brightest and/or dynamically youngest
satellites can be recovered.

Like the original GC3 method our extended version is limited to recovering
remnants of satellites of which the orbits are broadly confined to a plane. This
makes this method well suited for probing the outer halo of our Galaxy where the
potential is more nearly spherical or axisymmetric and where dynamical
timescales are long. The mGC3 method thus forms an important complement to the
phase space structure characterization methods which will be used in the inner
halo, where due to the short dynamical timescales phase space substructures are
harder to detect. In these regions it will be mandatory to use very accurate
measurements of integrals of motion employing methods such as those as proposed
in \cite{hel00} and \cite{FGomez010}. In both cases only the use of the highest
accuracy parallaxes \citep[relative errors better than about 10 per cent, see also][]{FGomez010B} allows the recovery of phase space substructure. In contrast
the mGC3 method works with $30$ per cent accurate parallax data making it suitable for
probing larger distances.

The Gaia mission is expected to map the immediate Solar neighbourhood to very
high accuracy resulting in some 10 million stars with distances known to better
than a few per cent. This will enable a much more accurate calibration of other
distance indicators, in particular photometric indicators, which can then be used
to extend the mGC3 method also to large samples beyond 40 kpc. In addition the
Gaia mission will result in a first order smooth dynamical model of the Milky
Way Galaxy which can be used to construct an accurate map of the background
mGC3 pole count map. Subtracting a more accurate background map will further
enhance the efficiency of the mGC3 method.

We note here two lines of investigation that should be pursued to further
investigate and enhance the mGC3 method. (i) The efficiency of the mGC3 method
should be explored more extensively by more widely sampling the satellite
orbital parameter space. More precise limits on the orbital morphologies and
dynamical ages, for which satellites remnants can still be recovered, can then
be obtained. In addition, the simulation of the satellites could be made more
sophisticated by following the suggestions made in the conclusions of B05.
(ii) In connection with these studies it is interesting to investigate to what
extent the requirement of single peaks in the pole count map for identifying
remnants can be relaxed. This would enable us to account for the precession of
the orbital angular momentum vector and thereby recover a larger fraction of
the stellar population of a given disrupted satellite. A hint of this
possibility can already be seen in the double peaked structure of the
signature of the satellite S2 in the map presented in Fig.\ref{fig:polemap_s2}.

Finally, the mGC3 method is not perfect and will certainly pick up stars from
the field population that appear to belong to a satellite because they are
measured to have a compatible orbital angular momentum vector. These should be
weeded out by making use of the fact that the different star formation histories
for the different satellites will lead to different abundance patterns in their
stellar populations. Gross distinctions (such as large overall metallicity
differences) can be made photometrically. However, eventually the stars
identified as part of a potential satellite remnant should be targeted for
detailed spectroscopic follow up in order to definitively `tag' the stars to their
progenitor galaxy and obtain accurate information on the progenitor's time of
accretion.

We look forward the time ahead of us when, thanks to the Gaia mission,
complementary spectroscopic data, and efficient methods of characterizing
substructure in phase space, we will enter the era of precision Galactic
archaeology.

We acknowledge support from CONACyT--M\'exico grant 60354. 
C. Mateu acknowledges support from the predoctoral grant of the 
Academia Nacional de Ciencias F\'{\i}sicas, Matem\'aticas y Naturales of Venezuela.

\appendix
\section{COORDINATE TRANSFORMATIONS}\label{s:appendix}

In the following we describe the coordinate transformation involved
in expressing Galactocentric position $\mathbf{r}_\mathrm{gal}$ and velocity $\mathbf{v}_\mathrm{gal}$ 
in terms of the heliocentric observable quantities $(l,b,\varpi,v_r,\mu_l,\mu_b)$,  
latitude, longitude, parallax, radial velocities and proper motions respectively.

\begin{figure} 
\includegraphics[width=84mm]{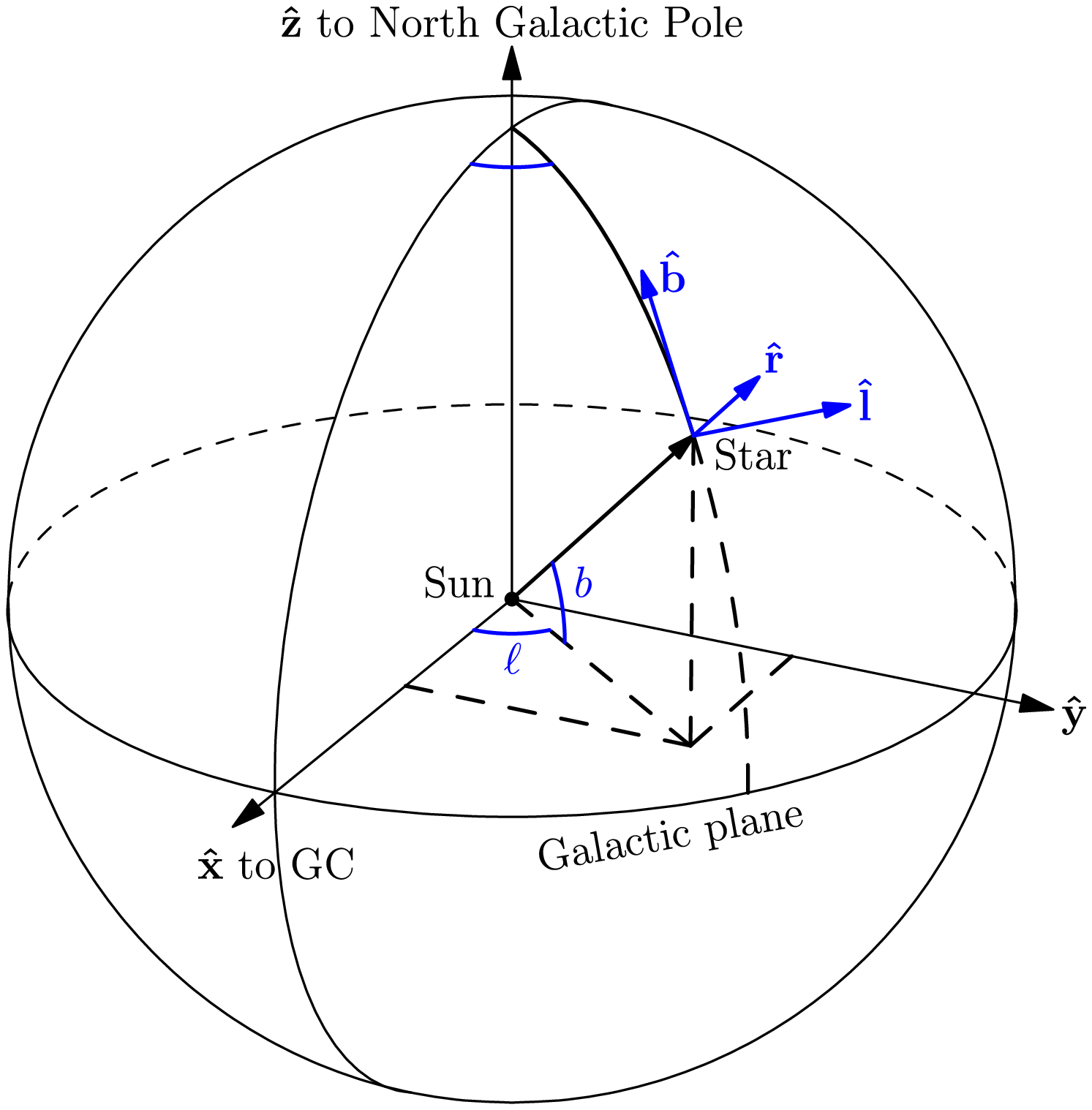} 
\caption{Relative orientation of the unit vectors $\lbrace \mathbf{\hat{r}},\mathbf{\hat{l}},\mathbf{\hat{b}} \rbrace$ and $\lbrace \mathbf{\hat{x}},\mathbf{\hat{y}},\mathbf{\hat{z}} \rbrace$, which respectively define the spherical heliocentric and cartesian Galactocentric reference frames used.}
\label{fig:axes}
\end{figure}

The Galactocentric position and velocity are expressed as

\begin{equation}
 \begin{array}{l}
  \mathbf{r}_\mathrm{gal} = \mathbf{r}_\odot + A_p\varpi^{-1}\mathbf{\hat{r}} \\
 \mathbf{v}_\mathrm{gal} =  \mathbf{v}_\odot + v_r \mathbf{\hat{r}} + \varpi^{-1} \big [ (A_v \mu_l \cos{b}) \mathbf{\hat{l}}+ (A_v\mu_b) \mathbf{\hat{b}} \big]
 \end{array} 
\end{equation}

where $A_p=1000$ mas$\cdot$pc, $A_v=4.74047$ yr \kms and $\lbrace \mathbf{\hat{r}},\mathbf{\hat{l}},\mathbf{\hat{b}} \rbrace$ are the unit vectors in a spherical heliocentric reference frame. The latter are expressed in terms of the unit vectors in the cartesian heliocentric reference frame $\lbrace \mathbf{\hat{x}},\mathbf{\hat{y}},\mathbf{\hat{z}} \rbrace$ as follows

\begin{equation}\label{AP1_rlb}
 \begin{array}{l}
  \mathbf{\hat{r}} = (\cos{l}\cos{b}) \mathbf{\hat{x}}+(\sin{l}\cos{b}) \mathbf{\hat{y}}+(\sin{b}) \mathbf{\hat{z}}\\
  \mathbf{\hat{l}} = -(\sin{l})\mathbf{\hat{x}}+(\cos{l})\mathbf{\hat{y}} \\
  \mathbf{\hat{b}} = -(\sin{b}\cos{l}) \mathbf{\hat{x}}-(\sin{b}\sin{l}) \mathbf{\hat{y}}+(\cos{b}) \mathbf{\hat{z}} 
 \end{array} 
\end{equation}

where $\mathbf{\hat{x}}$ points from the Sun in the GC-Sun direction, $\mathbf{\hat{y}}$ points
in the direction of Galactic rotation and $\mathbf{\hat{z}}$ points towards the North Galactic Pole.
The relative orientation of the unit vectors in both reference frames is shown in Figure \ref{fig:axes}.

Finally, $\mathbf{r}_\mathrm{gal}$ and $\mathbf{v}_\mathrm{gal}$ are

\[
 \begin{array}{ll}
  \mathbf{r}_\mathrm{gal} = & \mathbf{r}_\odot + A_p\varpi^{-1} [ (\cos{l}\cos{b}) \mathbf{\hat{x}}+(\sin{l}\cos{b}) \mathbf{\hat{y}}+(\sin{b}) \mathbf{\hat{z}} ] \\
 \mathbf{v}_\mathrm{gal} = & \mathbf{v}_\odot + [ v_r\sin{b} + A_v\varpi^{-1}\mu_b\cos{b}]\mathbf{\hat{z}}\\
& +[ v_r\cos{l}\cos{b} -A_v\varpi^{-1}(\mu_l\cos{b}\sin{l} +\mu_b\sin{b}\cos{l})]\mathbf{\hat{x}} \\ 
& +[ v_r\sin{l}\cos{b} + A_v\varpi^{-1}(\mu_l\cos{b}\cos{l} -\mu_b\sin{b}\sin{l})]\mathbf{\hat{y}} \\ 
 \end{array} 
\]

As can be seen from these equations,  $\mathbf{r}_\mathrm{gal}$ and $\mathbf{v}_\mathrm{gal}$ depend 
upon $(l,b)$ and the reciprocal of the parallax $\varpi$, as noted in Section \ref{s:mgc3_practical_impl}.

\end{document}